\documentclass[%
 reprint,
 amsmath,amssymb,
 aps,
]{revtex4-1}
\usepackage{graphicx} 
\usepackage{bm}
\usepackage[utf8]{inputenc}
\usepackage[T1]{fontenc}
\usepackage{mathptmx}
\usepackage{etoolbox}
\usepackage{amsmath}
\usepackage{amssymb}
\usepackage[colorlinks=true,linkcolor=blue]{hyperref}
\hypersetup{
  colorlinks = true,
  citecolor = blue,
  urlcolor = blue
}

\newcommand{\zpm}{\bm{z^{\pm}}}
\newcommand{\zmp}{\bm{z^\mp}}

\newcommand{\bb}{\bm{b}}
\newcommand{\bv}{\bm{v}}
\newcommand{\meanb}{\bm{b_0}}
\newcommand{\bnabla}{\bm{\nabla}}

\newcommand{\kparallel}{k_\parallel}

\begin{document}
\title{Universal relations between parallel and perpendicular spectral power law exponents in non-axisymmetric magnetohydrodynamic turbulence}
\author{Ramesh Sasmal}
\author{Supratik Banerjee}%
 \email{sbanerjee@iitk.ac.in}
\affiliation{ 
Department of Physics, Indian Institute of Technology Kanpur, Uttar Pradesh 
}%

\begin{abstract}
Following a general heuristic approach, algebraic constraints are established between the parallel and perpendicular power-law exponents of non-axisymmetric, highly aligned magnetohydrodynamic turbulence, both with and without strong imbalance between the Els\"asser variables. Such relations are universal both for the regimes of weak and strong turbulence and are useful to predict the corresponding turbulent power spectra. For scale-dependent alignment, a Boldyrev-type $k^{-3/2}$ perpendicular spectrum emerges transverse to the direction of alignment whereas a $k^{-5/3}$ spectrum is obtained for the same if the alignment becomes scale-independent. However, regardless of the nature of alignment, our analysis consistently yields a $k_{\parallel}^{-2}$ spectrum - commonly observed in both numerical simulations and \textit{in-situ} data of solar wind. In appropriate limit, previously obtained algebraic relations and power spectra for axisymmetric MHD turbulence (Galtier, Pouquet and Mangeney, Physics of Plasmas, 2005)  are successfully recovered. Finally, more realistic relations capturing weak Alfv\'en turbulence (with constant $k_{\parallel}$) and the transition to strong turbulence are derived along with their corresponding power spectra.             
\end{abstract}

\maketitle
\section{Introduction}
Turbulence, a highly nonlinear and multiscale phenomenon ubiquitous in nature, is necessary to understand the efficient mixing, structure formation, and rapid heating of the flow medium. A fluid flow becomes turbulent at a high Reynolds number with dominating nonlinearity over viscous effects. Both space and laboratory plasmas often exhibit turbulent behavior by means of fluctuations of different dynamical variables (velocity and magnetic fields). For fluctuations sufficiently larger than the ion inertial scale and ion gyroradius, plasma turbulence can be modeled  in the framework of ordinary magnetohydrodynamics (MHD).  In a fully developed MHD turbulence, the total energy (sum of kinetic and magnetic energies) nonlinearly cascades from the largest scale of energy injection to the smallest scales of dissipation. Unlike hydrodynamic turbulence where the mean velocity field can be eliminated by Galilean transformation, the effect of uniform background magnetic field $\bf B_0$ can not be eliminated from the MHD equations.  Consequently, the properties of MHD turbulence show notable variation along and perpendicular to ${\bf B}_0$. Whereas the fluctuations are nonlinearly deformed and eventually fragmented in the plane perpendicular to ${\bf B}_0$, they are practically advected along it. As $ B_0$ increases, the parallel advection becomes stronger than the perpendicular fluctuations  leading to $\ell_\parallel \gg \ell_\perp$ or equivalently $k_\perp \gg k_\parallel$ for the wave vector ${\bf k}= k_{\parallel} \hat{B}_0 + {\bf k}_\perp$ \citep{kadomtsev1973nonlinear, strauss1976nonlinear, montgomery1982, zank1992equations}. This eventually leads to a power anisotropy where the $E_\perp (k_\parallel, k_\perp) \neq E_\parallel (k_\parallel, k_\perp)$ \citep{horbury2012, matthaeus2012}. In addition, the parallel and perpendicular energy power spectra may also follow different power laws entailing spectral index anisotropy given by $E(k_\perp, k_\parallel)\sim k_\perp^{-\alpha_\perp}k_\parallel^{-\alpha_\parallel}$. 

Based on the mutual importance of linear and nonlinear dynamics in the energy cascade, a turbulence flow can be classified as weak or strong. A quantitative measure to determine this is given by the ratio $\chi = \tau_{lin}/\tau_{nl}$. For incompressible MHD turbulence, $\tau_{lin}$ is practically given by the Alfv\'en time $\tau_{A}\sim \ell_\parallel/V_A\sim (k_\parallel V_A)^{-1}$, where $V_A=B_0/\sqrt{\mu_0 \rho_0}$ is the Alfv\'en speed with  $\mu_0$ and $\rho_0$ being free space permittivity and the fluid density, respectively. On the other hand, the nonlinear time is given by $\tau_{nl}\sim (k_\perp v_\perp)^{-1}$ and hence we obtain 
\begin{equation}
    \chi=\frac{\tau_A}{\tau_{nl}}\sim \frac{k_\perp v_\perp}{k_\parallel V_A}. 
\end{equation}
A weak turbulence regime corresponds to $\chi \ll 1$ whereas a strong turbulence is characterized by $\chi \gtrsim 1$. Iroshnikov and Kraichnan (IK) were the first to predict the energy spectra of MHD turbulence in the weak regime \citep{iroshnikov1964turbulence, kraichnan1965inertial}. In the presence of a considerable $\bf B_0$, assuming an energy cascade due to the interactions of counter propagating  Alfv\'en wave packets, they proposed an isotropic energy spectra $E(k)\sim k^{-3/2}$.  Despite the presence of a strong $B_0$, the assumption of isotropy is not inconsistent in this case as a regime with $\chi\ll1$ remains possible even if $k_\perp \sim k_\parallel$. Nevertheless, a more realistic spectra can be obtained by incorporating anisotropy, leading to the corresponding energy spectra $E(k_\perp, k_\parallel)\sim k_\perp^{-2}k_\parallel^{-1/2}$ \citep{csNg1997, galtier2005spectral}. For strong turbulence regime, $\chi$ has to be $\gtrsim 1$. A strong $B_0$, therefore, requires the ratio $k_\perp/k_\parallel$ to be large, resulting in a prominent wave vector anisotropy.  Assuming a state of critical balance, where $\tau_A\sim \tau_{nl} \,\,(\chi\sim 1)$, Goldriech and Sridhar (GS) showed $k_\parallel\sim k_\perp^{2/3}$ and hence an anisotropic spectra $E(k_\perp, k_\parallel)\sim k_\perp^{-5/3}k_\parallel^{-1}$ \citep{goldreich1995}. This leads to the reduced energy spectra $E(k_\perp)\sim k_\perp^{-5/3}$ and  $E(k_\parallel)\sim  k_\parallel^{-2}$ which are observed both in numerical simulations \citep{ Cho_2000, Cho_2002, dmitruk2003energy, gomez2005parallel, dmitruk2005direct} and in-situ observations of solar wind turbulence \citep{matthaeus1982, horbury2008, deepali2021, saksee2022}. Assuming $\chi$ to be  constant, \citet{galtier2005spectral} proposed a general constitutive relation between  $\alpha_\parallel$ and $\alpha_\perp$ as $3\alpha_\perp+2\alpha_\parallel=7$, which is applicable for both strong and weak turbulence regimes. In particular, one can observe, for weak turbulence $\alpha_\perp=2$ and $\alpha_\parallel=1/2$ whereas $\alpha_\perp=5/3$ and $\alpha_\parallel=1$ for strong turbulence. 

Regardless of the success of critical balance theory, some high resolution numerical simulations of MHD turbulence reported an unusual -3/2 energy spectra which could not be explained in the framework proposed by GS  \citep{ Maron_2001, muller2003}. Relaxing the axisymmetric condition, Boldyrev proposed a plausible phenomenological solution to the aforementioned problem \citep{boldyrev2006}. In particular, he assumed a forced MHD turbulence with a high tendency of alignment between the velocity and magnetic field fluctuations \footnote{At this point, one should note that dynamic alignment in decaying MHD turbulence has a specific meaning of complete alignment of  the velocity and magnetic field fluctuations with approximately similar amplitude \textit{i.e.} $v\sim b$ \citep{banerjee2023universal}. Here, dynamic alignment merely means the small misalignment between the fields irrespective of their amplitudes. } and suggested the emergence of two distinct perpendicular length scales $\xi$ and $\lambda$, along and perpendicular to the alignment, respectively.  However, unlike a decaying MHD turbulence, perfect alignment (or anti-alignment) is not achieved in a forced turbulence and therefore, the velocity and magnetic field fluctuations subtend small angles  $\theta_\lambda$ in the plane perpendicular to $\bf B_0$ and $\tilde{\theta}_\lambda$ in the plane containing $\bf B_0$ and the magnetic field fluctuations. Finally, by minimizing the uncertainty of the resultant mismatch angle  $\sqrt{\theta_\lambda^2+\tilde{\theta}_\lambda^2}$, one obtains $\theta_\lambda\sim \tilde{\theta}_\lambda$ and consequently obtains the reduced energy spectra as $E(k_\lambda)\sim k_\lambda^{-3/2}$, $E(k_\xi)\sim k_\xi^{-5/3}$, and $E(k_\parallel)\sim k_\parallel^{-2}$. These reduced spectra can collectively be  presented  in terms of a the three dimensional  non-axisymmetric energy spectra $E(k_\lambda, k_\xi, k_\parallel)\sim k_\lambda^{-3/2} k_\xi^{-1}k_\parallel^{-1}$.
Whereas the initial studies assumed an approximate balance between the Els\"asser variables \textit{i.e.} $z^+\sim z^-$ \citep{Boldyrev_2005, boldyrev2006}, similar argument was extended later for imbalanced MHD case with $z^+\gg z^-$ where the power spectra of $z^+$ mimics the three-dimensional energy spectra obtained in balanced MHD turbulence \citep{boldyrev2009dynamic, podesta2010}.

The constitutive relation obtained by \citet{galtier2005spectral} assumes axisymmetry and hence consists of only one perpendicular power law exponent ($\alpha_\perp$) for the three dimensional energy spectra. However, the energy spectra obtained from Boldyrev's phenomenology is necessarily non-axisymmetric involving two perpendicular power law exponents ($\alpha_\lambda, \, \alpha_\xi$) and therefore, is expected to follow a more general version of the aforementioned constitutive relation. In addition, the said constitutive relation is established for balanced MHD turbulence only. Finally, the constitutive relation was derived for a constant value of $\chi$ and hence could capture neither the range of scales for weak Alfv\'enic turbulence where $k_\parallel$ remains constant and $\chi $ varies nor the transition between weak to strong turbulence \citep{meyrand2016}.

In this paper, following Boldyrev's phenomenology of alignment, we derive constitutive relations between $\alpha_\lambda, \, \alpha_\xi,$ and $\alpha_\parallel$  for three dimensional non-axisymmetric MHD turbulence. Assuming $\chi $ to be constant, we separately derive the constitutive relations corresponding to $E^+$ and $E^-$ (power spectra of $z^+$ and $z^-$, respectively) for both weak and strong imbalanced MHD turbulence with $z^+\gg z^-$. 
Under suitable conditions, we successfully recover the constitutive relation previously derived for the axisymmetric case. Finally, a general constitutive relation is obtained for constant $k_\parallel$ and varying $\chi$  capturing the regime of weak Alfv\'enic turbulence including the transition to the  regime of critical balance.

The paper is organized as follows: in Sec. \ref{model}, we present the governing equations and derive the non-axisymmetric constitutive relations for $E^+$ and $E^-$ in \ref{Eplus} and \ref{eminus}, respectively. For each of $E^+$ and $E^-$, we separately discuss the cases for strong and weak turbulence. In section \ref{varrying_chi}, we derive a general form of constitutive relation with varying $\chi$ and hence predict some constitutive relations for weak turbulence. Finally, in sec. \ref{discussion}, we summarize our results and conclude.

\section{Constitutive relations with constant $\chi$}\label{model}

In the presence of a uniform magnetic field, the governing equations of incompressible MHD are given by 
\small{
\begin{align}
    &\bnabla\cdot\bv=0, \\
    &\bnabla\cdot\bb=0,\\
    &\frac{\partial \bv}{\partial t}+(\bv\cdot\bnabla)\bv=-\bnabla p^*+ [(\meanb + \bb)\cdot\bnabla]\bb+\bm{f}+\nu\bnabla^2\bv,\\
    &\frac{\partial \bb}{\partial t}+(\bv\cdot\bnabla)\bb=[(\meanb +\bb)\cdot\bnabla]\bv+\eta\bnabla^2\bb,
\end{align}}

\noindent where $\meanb =\bm{B_0}/\sqrt{\mu_0}$ and $\bb$ represent the background magnetic field and the fluctuating magnetic field with respect to $\bf b_0$ in Alfv\'en units, respectively.  In terms of Els\"asser variables ($\zpm=\bv\pm\bb$), the above set of equations is written as
\small{
\begin{align}
    \bnabla\cdot\zpm&=0,\\
    \frac{\partial \zpm}{\partial t}\mp(\meanb \cdot\bnabla)\zpm&+(\zmp\cdot\bnabla)\zpm=-\bnabla p^*+\bm{f}+\bm{d_\pm}, 
\end{align}}

\noindent where $d_\pm=\bnabla^2(\nu^+\zpm+\nu^-\zmp)$ and $\nu^\pm=(\nu\pm\eta)/2$. 
Unlike the phenomenology proposed by IK and GS, the assumption of axisymmetry does not hold when there is a strong alignment between $\bv$ and $\bb$. In this situation, in addition to the parallel length scale $l$, we have two distinct perpendicular length scales $\xi$ and $\lambda$  along and perpendicular to the direction of alignment, respectively such that $l\gg\xi\gg\lambda$. In this case, the scale-dependent angle ($\theta_\lambda$) between $v_\lambda$ and $b_\lambda$ becomes very small. Further assuming $v_\lambda\sim b_\lambda$, one can express the Els{\"a}sser variable as $ z^+ \sim v_\lambda$ and $ z^-\sim  v_\lambda \theta_\lambda$, where $v_\lambda=\sqrt{<(\bv \cdot \hat{\lambda})^2>}$ and $b_\lambda=\sqrt{<(\bb \cdot\hat{\lambda})^2>}$ \citep{boldyrev2009dynamic, Perez2009} . Evidently, this is a case of strongly imbalanced turbulence as $ z^+\gg  z^-$ and one needs to obtain the spectral power laws for $E^+$ and $E^-$ separately. In case of balanced turbulence \citep{boldyrev2006}, the power spectra of $z^+$ becomes similar to $z^-$ and therefore the total energy spectra can be represented by either of them.  
\begin{figure}
    \centering
    \includegraphics[width=9.5 cm]{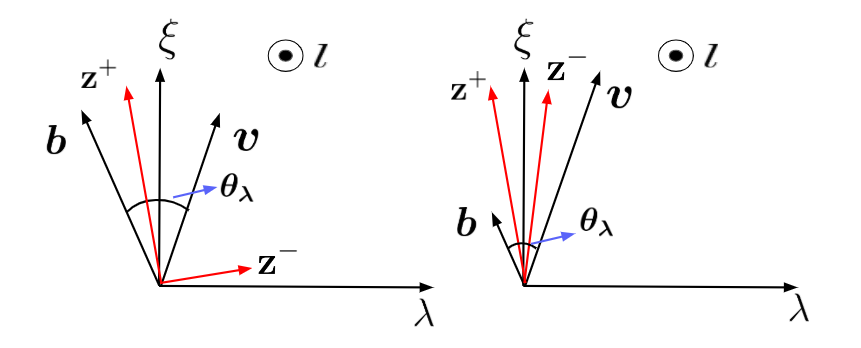}
    \caption{Imbalanced (left) and balanced  (right) MHD turbulence according to Boldyrev's phenomenology.}
    \label{fig1}
\end{figure}

\subsection{ Power spectra of $E^+$}\label{Eplus}

The rate of  scale-to-scale transfer of $(z^{+})^2$ is given by
\begin{equation}
    \varepsilon^{+}\sim \frac{( z^+)^2}{\tau_{tr}^+}\sim \frac{v_\lambda^2}{\tau_{tr}^+}, \label{epsilon}
\end{equation}
where $\tau_{tr}^+$ is the average transfer time. For strong turbulence $\tau_{tr}^+\sim \tau_{nl}^+$ whereas for weak turbulence involving $n$-wave interaction, $\tau_{tr}^+\sim (\tau_{nl}^+)^{2n-4}/\tau_A^{2n-5}$, where $\tau_A\sim l/b_0$ \citep{Galtier2022}. Since Alfv\'en wave turbulence is dominated by three-wave interactions, we have $n=3$ and hence  $\tau_{tr}^+\sim (\tau_{nl}^+)^{2}/\tau_A$. Defining, $\chi^+=\tau_A/\tau_{nl}^+$, one can also have the transfer time as $\tau_{tr}^+\sim \tau_{nl}^+/(\chi^+)^{\beta}$ with $\beta=0$ and $1$ corresponding to the cases of strong and Alfv\'en wave turbulence, respectively. Furthermore, the nonlinear interactions occur only between counter-propagating Alfv\'en wave packets \citep{Biskamp_2003, banerjee_thesis, Galtier2022} and hence the nonlinear time for $E^+$ is given by $\tau_{nl}^+\sim \lambda/ z^-\sim \lambda/(v_\lambda \theta_\lambda)$, as $z^-$ mainly propagates along $\lambda$ (see figure (\ref{fig1})). From definition, one can write
\begin{align}
    \chi^+=\frac{\tau_A}{\tau_{nl}^+}\sim \frac{v_\lambda \theta_\lambda l}{\lambda b_0}
    \Rightarrow l\sim \chi^+\frac{\lambda b_0}{v_\lambda \theta_\lambda },\label{epsilon3}
\end{align}

\noindent and  the respective transfer time can then be written as
\begin{equation}
    \tau_{tr}^+\sim \frac{(\tau_{nl}^+)^{(\beta+1)}}{\tau_A^\beta} \sim \left(\frac{\lambda}{v_\lambda \theta_\lambda}\right)^{\beta+1}\left(\frac{b_0}{l}\right)^{\beta}. \label{epsilon4}
\end{equation}

\noindent Eliminating $l$  by using the relation (\ref{epsilon3}) in Eq. (\ref{epsilon4}), we can get
\begin{equation}
    \tau_{tr}^+\sim \frac{\lambda}{v_\lambda \theta_\lambda (\chi^+)^{\beta}}.
\end{equation} 
\noindent Finally, substituting the above relation in Eq. (\ref{epsilon}), we get
\begin{equation}
    \varepsilon^+ \sim \frac{v_\lambda^3\theta_\lambda (\chi^+)^{\beta}}{\lambda}
  \Rightarrow v_\lambda\sim \left(\frac{\lambda}{\theta_\lambda}\right)^{\frac{1}{3}}\left(\frac{ \varepsilon^+ }{(\chi^+)^{\beta}}\right)^{\frac{1}{3}}.\label{vlambda}
\end{equation}

\noindent  The nonlinear time scale $\tau_{nl}^+$ can equivalently be estimated through the deformation of $z^+$ along its own direction of propagation as
\begin{align}
    \tau_{nl}^+\sim \frac{\xi}{z^+} \sim \frac{\xi}{v_\lambda}\Rightarrow \xi\sim v_\lambda\tau_{nl}^+\sim \frac{\lambda}{\theta_\lambda}\Rightarrow \frac{\lambda}{\xi}\sim\theta_\lambda \label{epsilon7}.
\end{align}

\noindent Following \citet{boldyrev2006}, the scale-dependence of $\theta_\lambda$ can be written as $\theta_\lambda\sim\lambda^{\mu/(3+\mu)}$ where $\mu$ is a non-negative constant ascertaining $\theta_\lambda$ to be small. Assuming $\chi^+$ and $\varepsilon^+$ to be constant, the angle of mismatch in the vertical plane (containing $\meanb$ and $\bb$) is given by
\begin{equation}
    \tilde{\theta}_\lambda\sim\frac{\xi}{l}\sim \frac{v_\lambda }{b_0\chi^+}\sim \lambda^{\frac{1}{\mu+3}}.
\end{equation}

\noindent  The total angle of mismatch between the fluctuations is, therefore,  given by $\phi_\lambda=\sqrt{\theta_\lambda^2+\tilde{\theta}_\lambda^2}$ and the `uncertainty' of $\phi_\lambda$ is minimum when $\theta_\lambda\sim \tilde{\theta}_\lambda$ implying $\mu=1$. So, in the perpendicular plane, the scale-dependent angle between the fluctuations scales as $\theta_\lambda\sim \lambda^{1/4}$ \citep{boldyrev2009dynamic, Perez2009}.  However, for the sake of generality, we keep $\theta_\lambda\sim \lambda^\alpha$ for the rest of the calculation allowing the analysis to hold for any other phenomenology of non-axisymmetric turbulence with  small $\theta_\lambda$ but $\alpha \neq 1/4$. Incorporating the scale-dependence of $\theta_\lambda$ in Eq. (\ref{vlambda}), one obtains
\begin{equation}
    v_\lambda\sim \lambda^{\frac{(1-\alpha)}{3}}\left(\frac{ \varepsilon^+ }{(\chi^+)^{\beta}}\right)^{\frac{1}{3}},\label{epsilon9}
\end{equation}

\noindent putting the above expression in relation (\ref{epsilon3}), the parallel length scale can be expressed as 
\begin{equation}
    l\sim b_0\frac{(\chi^+)^{(1+\frac{\beta}{3})}}{(\varepsilon^+)^{\frac{1}{3}}}\lambda^{\frac{2}{3}(1-\alpha)}\nonumber.
\end{equation}
 \noindent Given the scale-independence of $b_0$, $\chi$, and  $\varepsilon^+$, we obtain the relation between the three length scales as
 \begin{align}
     l &\sim \lambda^{\frac{2}{3}(1-\alpha)} \Rightarrow k_\parallel\sim k_\lambda^{\frac{2}{3}(1-\alpha)}\label{epsilon10},\\
     \xi&\sim \lambda^{1-\alpha}\Rightarrow k_\xi\sim k_\lambda^{1-\alpha}.\label{epsilon11}
 \end{align}
 
\noindent By definition, $(z^+)^2\sim v_\lambda^2\sim E^+(k_\lambda, k_\xi, k_\parallel)k_\lambda k_\xi k_\parallel $, where $E^+(k_\lambda, k_\xi, k_\parallel)$ represents three-dimensionally non-axisymmetric power spectra for $z^+$. Assuming $E^+(k_\lambda, k_\xi, k_\parallel)\sim k_\lambda^{-\alpha_\lambda}k_\xi^{-\alpha_\xi} k_\parallel^{-\alpha_\parallel}$, where $\alpha_\lambda$, $\alpha_\xi$, and $\alpha_\parallel$ represent the power law indices for the length scales $\lambda$, $\xi$, and $l$, respectively, we obtain
\begin{equation}
    E^+(k_\lambda, k_\xi, k_\parallel)k_\lambda k_\xi k_\parallel \sim k_\lambda^{\frac{2}{3}(\alpha-1)}.\label{epsilon13} 
\end{equation}

\noindent Finally using Eq. (\ref{epsilon10}) and (\ref{epsilon11}), we derive the general constitutive relation as

{\small
\begin{equation}
     3\alpha_\lambda+3(1-\alpha)\alpha_\xi+2(1-\alpha)\alpha_\parallel=10-7\alpha.\label{epsilon14}
\end{equation}}

\noindent For Boldyrev's phenomenology $\alpha= 1/4$, the general relation becomes
\begin{equation}
    4\alpha_\lambda+3\alpha_\xi+2\alpha_\parallel=11.\label{epsilon15}
\end{equation}

\noindent For strong turbulence, $\tau_{tr}^+\sim\tau_{nl}^+\sim \lambda/v_\lambda \theta_\lambda$ and therefore,
\begin{align}
    &\varepsilon^+\sim \frac{(z^+)^{2}}{\tau_{tr}^+}\sim \frac{v_\lambda^3\theta_\lambda}{\lambda}\sim \frac{v_\lambda^3}{\lambda^{\frac{3}{4}}}\label{epsilon+}\\
    &\Rightarrow v_\lambda^2\sim \lambda^{\frac{1}{2}}\sim k_\lambda^{-\frac{1}{2}}.
\end{align}

\noindent Again for non-axisymmetric power spectra of $z^+$, we get
\begin{align}
    &E^+(k_\lambda, k_\xi, k_\parallel)k_\lambda k_\xi k_\parallel\sim (z^+)^{2}\sim v_\lambda^2\sim k_\lambda^{-\frac{1}{2}}\nonumber\\
    \Rightarrow& E^+(k_\lambda, k_\xi, k_\parallel)\sim k_\lambda^{-\frac{3}{2}}k_\xi^{-1}k_\parallel^{-1}.
\end{align}

\noindent The above spectra corresponds to $\alpha_\lambda=3/2$, $\alpha_\xi=1$, and $\alpha_\parallel=1$ (also obtained from the exact relation of reduced magnetohydrodynamics \citep{sasmal2025}), which satisfy the constitutive relation (\ref{epsilon15}). Note that the above choice is not unique. Using an equivalent expression of $\tau_{nl}\sim \xi/v_\lambda$, one can obtain an energy spectra $E^+(k_\lambda, k_\xi, k_\parallel)\sim k_\lambda^{-1} k_\xi^{-5/3} k_\parallel^{-1}$, also satisfying the relation (\ref{epsilon15}). As mentioned previously, the transfer time scale corresponding to the weak transfer of $E^+ $ is given by 
\begin{equation}
    \tau_{tr}^+\sim \frac{(\tau_{nl}^+)^{2}}{\tau_A} \sim \left(\frac{\lambda}{v_\lambda \theta_\lambda}\right)^2\left(\frac{b_0}{l}\right).\label{equn24}
\end{equation}

\noindent The turbulent transfer rate, in this case, will, therefore, be 
\begin{equation}
    \varepsilon^+\sim\frac{(z^+)^{2}}{\tau_{tr}^+}\sim \frac{v_\lambda^4 \theta_\lambda^2 \,l}{\lambda^2 b_0 }
    \Rightarrow v_\lambda^2\sim \lambda^{\frac{3}{4}}l^{-\frac{1}{2}}\sim k_\lambda^{-\frac{3}{4}}\kparallel^{\frac{1}{2}}.\label{epsilon17}
\end{equation}

\noindent Finally, for the three-dimensional non-axisymmetric energy spectra of $z^+$, we get
\begin{align}
    &E^+(k_\lambda, k_\xi, k_\parallel)k_\lambda k_\xi k_\parallel\sim k_\lambda^{-\frac{3}{4}}\kparallel^{\frac{1}{2}}\nonumber\\
    \Rightarrow&E^+(k_\lambda, k_\xi, k_\parallel)\sim k_\lambda^{-\frac{7}{4}}k_\xi^{-1}\kparallel^{-\frac{1}{2}}. \label{epsilon18}
\end{align}

\noindent It is straightforward to verify the exponents of the above energy spectra also satisfy the relation (\ref{epsilon15}). For both the cases, the reduced spectra are given as $E^+(k_\lambda)\sim k_\lambda^{-3/2}$, $E^+(k_\xi)\sim k_\xi^{-5/3}$, and $E^+(k_\parallel)\sim k_\parallel^{-2}$.

\subsection{Power spectra of $E^-$}\label{eminus}

In this section, we are interested in the turbulent transfer of $(z^{-})^2$ with nonlinear time $\tau_{nl}^-\sim \xi/z^+\sim \lambda/(v_\lambda \theta_\lambda)$. Since non-linear timescale is the same for both the cases, the ratio $\chi^-\sim (v_\lambda \theta_\lambda l)/(\lambda b_0) \sim \chi^+$ will be same for both cases. The rate of turbulent transfer is, therefore, given by
\begin{align}
    \varepsilon^- &\sim \frac{(z^-)^{2}}{\tau_{tr}^-}\sim \frac{v_\lambda^3\theta_\lambda^3(\chi^-)^\beta}{\lambda}\label{epsilon-}\\
    \Rightarrow v_\lambda &\sim \frac{\lambda^{\frac{1}{3}}}{\theta_\lambda}\left(\frac{\varepsilon^-}{(\chi^-)^\beta}\right)^{\frac{1}{3}}\sim \frac{\lambda^{\frac{1}{3}}}{\theta_\lambda} ,\label{pseudo2}
\end{align}

\noindent where $\tau_{tr}^-\sim \tau_{nl}^-/(\chi^-)^\beta$ and both $\varepsilon^-$ and $\chi^-$ are assumed to be constant across scales. Similar as above, $\lambda/\xi\sim\theta_\lambda$ and taking $\theta_\lambda\sim \lambda^\alpha$, the mismatch in the vertical plane is given by
\begin{equation}
    \tilde{\theta}_\lambda\sim \frac{\xi}{l} \sim \frac{v_\lambda}{b_0 \chi^-}\sim \frac{\lambda^{\frac{1}{3}}}{\theta_\lambda}\sim \lambda^{\frac{1}{3}-\alpha}.
\end{equation}

\noindent The mismatch will be minimized when $\theta\sim \tilde{\theta}_\lambda$ which gives $\alpha=1/6$. Assuming $\theta_\lambda\sim \lambda^\alpha$, the general expression for cascade rate is given by  
\begin{align}
    \varepsilon^-&\sim \frac{(z^-)^{2}}{\tau_{tr}^-}\sim\frac{v_\lambda^3\theta_\lambda^3(\chi^-)^\beta}{\lambda}\\
    \Rightarrow v_\lambda&\sim \lambda^{\frac{(1-3\alpha)}{3}}\left(\frac{\varepsilon^-}{(\chi^-)^\beta}\right)^{\frac{1}{3}}.\label{e-}
\end{align}

\noindent Using the relation (\ref{e-}) in the expression of $\chi^-$, we get 
\begin{equation}
    l\sim \lambda^{\frac{2}{3}}b_0\frac{(\chi^-)^{(1+\frac{\beta}{3})}}{(\varepsilon^-)^{\frac{1}{3}}}\Rightarrow k_\parallel \sim k_\lambda^{\frac{2}{3}}.\label{parallel}
\end{equation}

\noindent The other perpendicular scale is given by $k_\xi\sim k_\lambda^{1-\alpha}$. Assuming three-dimensional anisotropic spectra $E^-(k_\lambda, k_\xi, k_\parallel)\sim k_\lambda^{-\alpha_\lambda}k_\xi^{-\alpha_\xi} k_\parallel^{-\alpha_\parallel}$, one can have $E^-(k_\lambda, k_\xi, k_\parallel)k_\lambda k_\xi k_\parallel\sim (z^-)^{2}\sim v_\lambda^2\theta_\lambda^2$. From Eq. (\ref{e-}), one gets
\begin{equation}
    E^-(k_\lambda, k_\xi, k_\parallel)k_\lambda k_\xi k_\parallel\sim v_\lambda^2\theta_\lambda^2\sim k_\lambda^{-\frac{2}{3}}.
\end{equation} 

\noindent Finally, comparing the above relation with Eq. (\ref{parallel}), we get
\begin{equation}
    3\alpha_\lambda+3(1-\alpha)\alpha_\xi+2\alpha_\parallel=10-3\alpha \label{relation2}.
\end{equation}

\noindent For $\alpha=1/6$, the constitutive relation reduces to 
\begin{equation}
    6\alpha_\lambda+5\alpha_\xi+4\alpha_\parallel=19\label{relation3}.
\end{equation}

\noindent For strong turbulence, $\tau_{tr}^-\sim \tau_{nl}^-\sim \lambda/v_\lambda \theta _\lambda$ and the corresponding transfer rate is given by
\begin{equation}
    \varepsilon^-\sim \frac{z^{-2}}{\tau_{tr}}\sim \frac{v_\lambda^3\theta_\lambda^3}{\lambda}\sim \frac{v_\lambda^3}{\lambda^{\frac{1}{2}}}
    \Rightarrow v_\lambda^2\sim \varepsilon^{-\frac{2}{3}}\lambda^{\frac{1}{3}}\sim k_\lambda^{-\frac{1}{3}}.
\end{equation}

\noindent The associated three-dimensional, non-axisymmetric spectra is, therefore, obtained as
\begin{equation}
    E^-(k_\lambda, k_\xi, \kparallel)\sim (z^-)^2 k_\lambda^{-1}k_\xi^{-1}k_\parallel^{-1}\sim k_\lambda^{-\frac{5}{3}}k_\xi^{-1}k_\parallel^{-1},
\end{equation}

\noindent where it is straightforward to check that the power law indices satisfy the constitutive relation (\ref{relation3}). For the weak transfer,
\begin{equation}
    \tau^-_{tr} \sim \frac{(\tau^-_{nl})^2}{\tau_A}\sim \frac{\lambda^2 b_0}{v_\lambda^2 \theta_\lambda^2 l},\label{equn38}
\end{equation}

\noindent and the cascade rate of $(z^-)^2$ can be obtained as
\begin{equation}
    \varepsilon^-\sim \frac{(z^-)^2}{\tau_{tr}}\sim \frac{v_\lambda^4 \theta_\lambda^4 l}{b_0 \lambda^2}
    \Rightarrow (z^-)^2\sim v_\lambda^2\theta_\lambda^2 \sim (\varepsilon b_0)^{\frac{1}{2}}  k_\lambda^{-1}k_\parallel^{\frac{1}{2}}.
\end{equation}

\noindent The corresponding three-dimensional, non-axisymmetric energy spectra is finally given by
\begin{equation}
    E^-(k_\lambda, k_\xi, \kparallel)\sim(z^-)^2 k_\lambda^{-1} k_\xi^{-1}k_\parallel^{-1}\sim k_\lambda^{-2}k_\xi^{-1}k_\parallel^{-\frac{1}{2}},
\end{equation}

\noindent where the exponents of the above energy spectra satisfy the general constitutive relation (\ref{relation3}). For both the cases, corresponding reduced power spectra are given by $E^-(k_\lambda)\sim k_\lambda^{-5/3}$, $E^-( k_\xi)\sim k_\xi^{-9/5} $, and $E^-(k_\parallel)\sim k_\parallel^{-2}$.

\subsection{Scale-independent imbalance and balanced MHD}\label{balanced}
 In the analysis above, we found different scale-dependence of $\theta_\lambda$ corresponding to the universal cascades of $(z^+)^2$ and $(z^-)^2$, separately. This can be understood from Eqs. (\ref{vlambda}) and (\ref{epsilon-}), implying the ratio $\varepsilon^-/\varepsilon^+\sim (\lambda/\xi)^2$ to be scale-dependent and hence a simultaneous scale-invariance of $\varepsilon^-$ and $\varepsilon^+$ is not guaranteed. Nevertheless, the invariance of both the cascade rates across the inertial scales can still be assured if the ratio $\lambda/\xi$ becomes scale-independent (but remains small). In such a case, $\alpha=0$ and hence both general constitutive relations reduce to 
\begin{equation}
    3\alpha_\lambda+3\alpha_\xi+2\alpha_\parallel=10.\label{alpha0}
\end{equation}

\noindent For strong and weak turbulence, the corresponding three-dimensional spectra are given by $E^\pm\sim k_\lambda^{-5/3}k_\xi^{-1}k_\parallel^{-1}$ and $E^\pm\sim k_\lambda^{-2}k_\xi^{-1}k_\parallel^{-1/2}$, respectively. For both cases, the reduced spectra are given by $E^\pm(k_\lambda)\sim k_\lambda^{-5/3}$, $E^\pm( k_\xi)\sim k_\xi^{-5/3} $, and $E^\pm(k_\parallel)\sim k_\parallel^{-2}$.

We now consider the case $v_\lambda\gg b_\lambda$ leading to balanced MHD case with $z^+\sim z^-\sim z_\lambda\sim v_\lambda$ for arbitrary angles between $\bv$ and $\bb$ (see figure (\ref{fig1})). The rate of energy transfer is given by
\begin{equation}
    \varepsilon \sim \frac{z^{2}_\lambda}{\tau_{tr}}\sim \frac{v_\lambda^{2}}{\tau_{tr}},
\end{equation}

\noindent where $\tau_{tr}$ is the transfer time scale. In this case, $\tau_{nl}\sim \xi/z_\lambda$ and $\tau_{tr}\sim \tau_{nl}/\chi^\beta$, where $\beta=0$ and $1$ corresponds to strong and weak turbulences, respectively and therefore, the cascade rate 
\begin{equation}
    \varepsilon \sim \frac{v_\lambda^3 \chi^{\beta}}{\xi}
  \Rightarrow v_\lambda^2\sim \xi^{\frac{2}{3}}\left(\frac{ \varepsilon }{\chi^{\beta}}\right)^{\frac{2}{3}}. \label{balance1}
\end{equation}

\noindent  This leads to an energy spectra  $E(k_\lambda, k_\xi, k_\parallel)\sim k_\lambda^{-1}k_\xi^{-5/3}k_\parallel^{-1}$. However, to obtain the directional energy spectra for $\lambda$, one needs to know the ratio $\lambda/\xi$ which explicitly depends on the phenomenology. Boldyrev's phenomenology \citep{boldyrev2006} corresponds to $\lambda\ll \xi$ and the ratio $\lambda/\xi\sim \theta_\lambda\sim \lambda^{1/4}$, leading to the general relation 
\begin{equation}
    4\alpha_\lambda+3\alpha_\xi+2\alpha_\parallel=11.\label{balance7}
\end{equation}

\noindent If the ratio $\lambda/\xi$ is scale-independent, the general relation becomes similar to Eq. (\ref{alpha0}). In particular, for axisymmetric turbulence, the ratio $\lambda/\xi\sim 1\Rightarrow k_\lambda\sim k_\xi\sim k_\perp$.
Assuming axisymmetric energy spectra $E(k_\perp, k_\parallel)\sim k_\perp^{-\alpha_\perp}k_\parallel^{-\alpha_\parallel}$, one can have 
\begin{equation}
   v_\lambda^2 \sim k_\perp^{-\alpha_\perp}k_\parallel^{-\alpha_\parallel}k_\perp k_\parallel\sim k_\lambda^{-\alpha_\lambda}k_\xi^{-\alpha_\xi}k_\parallel^{-\alpha_\parallel}k_\lambda k_\xi k_\parallel,
\end{equation}

\noindent and substituting $k_\lambda\sim k_\xi\sim k_\perp$ in the above relation, we obtain $\alpha_\perp=\alpha_\lambda+\alpha_\xi-1$. Therefore, the general relation (\ref{alpha0}) reduces to 
\begin{equation}
   3\alpha_\perp+2\alpha_\parallel=7\label{balance8},
\end{equation}

\noindent which is similar to the relation obtained previously \citep{galtier2005spectral}.

\section{Cases with variable $\chi$: transition from weak to strong turbulence}\label{varrying_chi}

The derivation of the relations (\ref{epsilon15}) and (\ref{relation3}) assumes constant $\chi^\pm$ across the scales. However, in weak Alfv\'enic turbulence, there is no turbulent transfer of energy along $\meanb$ and hence $k_\parallel$ becomes practically constant \citep{Matthaeus_Montgomery_1983, meyrand2016}. In this situation,  constant $\chi^\pm$ imply  $ {\lambda}\sim v_\lambda {\theta_\lambda} $, leading to $\tau_{tr}^+\sim \tau^-_{tr}$  (see equations (\ref{equn24}) and (\ref{equn38})) and the corresponding transfer rates are given by $\varepsilon^+\sim \lambda^2/\theta_\lambda^2 $, and $ \varepsilon^-\sim \lambda^2$. This is in clear contradiction to simultaneous cascades with scale-independent $\varepsilon^+$ and $\varepsilon^-$. To capture the entire spectrum of weak and strong turbulence under a universal theoretical framework, it is therefore reasonable to consider $\chi^\pm$ to vary across scales and $k_\parallel$ to be related to $k_\lambda$ as $k_\parallel\sim k_\lambda^{f(k_\lambda, k_\parallel)}$ where $f(k_\lambda, k_\parallel)=0$ represents the case of weak Alfv\'en wave turbulence. Assuming $\theta_\lambda\sim \lambda^\alpha$, the transfer time in the cascade for $(z^+)^2$ is therefore given by
\begin{equation}
    \tau^+_{tr}\sim \frac{\tau^+_{nl}}{(\chi^+)^\beta}\sim \left(\frac{\lambda^{1-\alpha}}{v_\lambda }\right)^{\beta+1}\left(\frac{b_0}{l}\right)^\beta,
\end{equation}

\noindent and the corresponding cascade rate becomes
\begin{align}
    \varepsilon^+&\sim \frac{(z^+)^{2}}{\tau^+_{tr}}\sim \frac{v_\lambda^{\beta+3}}{\lambda^{(1-\alpha)(\beta+1)}}\left(\frac{l}{b_0}\right)^\beta\nonumber\\
    \Rightarrow v_\lambda &\sim (\varepsilon^+b_0^\beta)^{\frac{1}{(\beta+3)}} k_\lambda^{\frac{(\alpha-1)(1+\beta)}{(\beta+3)}}
    k_\parallel^{\frac{\beta}{(\beta+3)}}.\label{varry}
\end{align}

\noindent Assuming anisotropic spectra $E^+(k_\lambda, k_\xi, k_\parallel)\sim k_\lambda^{-\alpha_\lambda}k_\xi^{-\alpha_\xi}k_\parallel^{-\alpha_\parallel}$ and following the similar mathematical steps as of the above sections, we get the general relation
\small{
\begin{align}
    \alpha_\lambda+(1-\alpha)\alpha_\xi +f(k_\lambda, k_\parallel)\alpha_\parallel=&(2-\alpha)-\frac{2(\alpha-1)(\beta+1)}{(\beta+3)}\nonumber\\
    &+f(k_\lambda, k_\parallel)\left(\frac{3-\beta}{\beta+3}\right).\label{parallel2}
\end{align}}

\noindent Employing similar analytical steps for the transfer of $(z^-)^2$, we also obtain the second general relation as
\begin{align}
    \alpha_\lambda+(1-\alpha)\alpha_\xi+f(k_\lambda, k_\parallel)\alpha_\parallel= &(2+\alpha)-\frac{2(\alpha-1)(\beta+1)-4\alpha}{(\beta+3)}\nonumber\\
    &+f(k_\lambda, k_\parallel)\left(\frac{3-\beta}{\beta+3}\right).\label{parallel3}
\end{align}

 \paragraph* {\hspace{-0.5cm}}{\textbf{i. Boldyrev's phenomenology:}}
 In weak Alfv\'enic turbulence, $\beta=1$ and $f(k_\lambda, k_\parallel)=0$. Furthermore, $k_\parallel$ does not change across scales and hence $\tilde{\theta}_\lambda\sim k_\parallel/k_\xi\sim 1/k_\xi\sim k_\lambda^{\alpha-1}$. The minimization of the total angle of mismatch gives $\alpha=1/2$ for both $z^+$ and $z^-$. Putting all these values in equations (\ref{parallel2}), for the transfer of $(z^+)^2$, we get
 \begin{equation}
     2\alpha_\lambda+\alpha_\xi=4.\label{51}
 \end{equation}
 
\noindent From Eq. (\ref{varry}), one obtains $v_\lambda^2\sim k_\lambda^{-1/2}$. Consequently, the energy spectra corresponds to $\alpha_\lambda=3/2$ and $\alpha_\xi=1$, satisfying the Eq. (\ref{51}). Similarly, for the weak cascade of $(z^-)^2$, the Eq. (\ref{parallel3}) reduces to 
\begin{equation}
    2\alpha_\lambda+\alpha_\xi=5.\label{52}
\end{equation}

\noindent In this case, $(z^-)^2\sim v_\lambda^2\theta_\lambda^2\sim k_\lambda^{-1}$ and the spectra corresponds to $\alpha_\lambda=2$ and $\alpha_\xi=1$, satisfying the Eq. (\ref{52}). For the case of Alfv\'enic weak turbulence, one can also verify $n^+_\lambda+n^-_\lambda= 7/2 $, where $n^+_\lambda$ and $n^-_\lambda$ are the exponents of $k_\lambda$ in $E^+(k_\lambda)$ and $E^-(k_\lambda)$, respectively. Similar relation can also be derived with respect to the reduced spectra along $k_\xi$. However, since $k_\lambda\gg k_\xi$, the perpendicular cascade is practically determined by the cascade along $k_\lambda$.

Beyond Alfv\'enic wave turbulence, $k_\parallel$ is no longer constant. For approximately constant $\chi^+$, we have $\alpha=1/4$ and $f(k_\lambda, k_\parallel)= 1/2$ for the transfer of $(z^+)^2$ and the Eq. (\ref{parallel2}) subsequently reduces to Eq. (\ref{epsilon15}). Similarly, for the turbulent transfer of $(z^-)^2$ with constant $\chi^-$, we obtain $\alpha=1/6$ and $f(k_\lambda, k_\parallel)=2/3$ reducing the Eq. (\ref{parallel3}) to the relation (\ref{relation3}). It is interesting to note, for the case with constant $\chi^\pm$, the general relations (\ref{parallel2}) and (\ref{parallel3}) become independent of $\beta$ and hence the power law indices $\alpha_\lambda$, $\alpha_\xi$, and $\alpha_\parallel$ satisfy the same relation for both weak and strong turbulent regimes.

\paragraph*{\hspace{-0.5cm}}{\textbf{ii. Scale-independent imbalance and axisymmetric turbulence:}}
In Boldyrev's phenomenology, a measure of imbalance is given by $z^-_\lambda/z^+_\lambda\sim \theta_\lambda$. If the imbalance is scale-independent, $\theta_\lambda$ is constant and hence $\alpha=0$, reducing both the relations (\ref{parallel2}) and (\ref{parallel3}) to
\begin{equation}
    \alpha_\lambda+\alpha_\xi+ f(k_\lambda, k_\parallel)\alpha_\parallel= 2+\frac{2(1+\beta)}{(\beta+3)}+f(k_\lambda, k_\parallel)\left(\frac{3-\beta}{\beta+3}\right).\label{parallel4}
\end{equation}

\noindent For weak Alfv\'en wave turbulence, $f(k_\lambda, k_\parallel)=0$ and $\beta=1$. Therefore, the general relation (\ref{parallel4}) becomes $\alpha_\lambda+\alpha_\xi=3$ which again reduces to $\alpha_\perp=2$ in the axisymmetric turbulence. The corresponding spectra, in this case, are given by $E^+(k_\perp)\sim k_\perp^{-2}$ and $E^-(k_\perp)\sim k_\perp^{-2}$ which makes $n^++n^-=4$, where $n^+$ and $n^-$ are the exponents of $k_\perp$ in $E^+$ and $E^-$, respectively. 

For strong turbulence, $\beta=0$ and therefore, the relation (\ref{parallel4}) becomes 
\begin{equation}
    \alpha_\lambda+\alpha_\xi+f(k_\lambda, k_\parallel)\alpha_\parallel= \frac{8}{3}+f(k_\lambda, k_\parallel).\label{parallel5}
\end{equation}

\noindent As it is evident, $f(k_\lambda, k_\parallel)=2/3$ corresponds to constant $\chi^\pm$, where the above relation reduces to $3\alpha_\lambda+3\alpha_\xi+2\alpha_\parallel=10$ which is identical to the relation (\ref{alpha0}). Finally, one can recover Eq. (\ref{balance8}) by incorporating additional assumption of axisymmetry. For balanced MHD,
 $z^+\sim z^-$ and the power spectra of total energy are similar to those obtained for $z^+$ in imbalanced MHD. Therefore, the corresponding weak Boldyrev's spectra are given by $E^\pm(k_\lambda, k_\xi)\sim k_\lambda^{-3/2}k_\xi^{-1}$ leading to $n_\lambda^++n_\lambda^-=3$.

\begin{figure*}
    \centering
    \includegraphics[width=14 cm, height=20 cm]{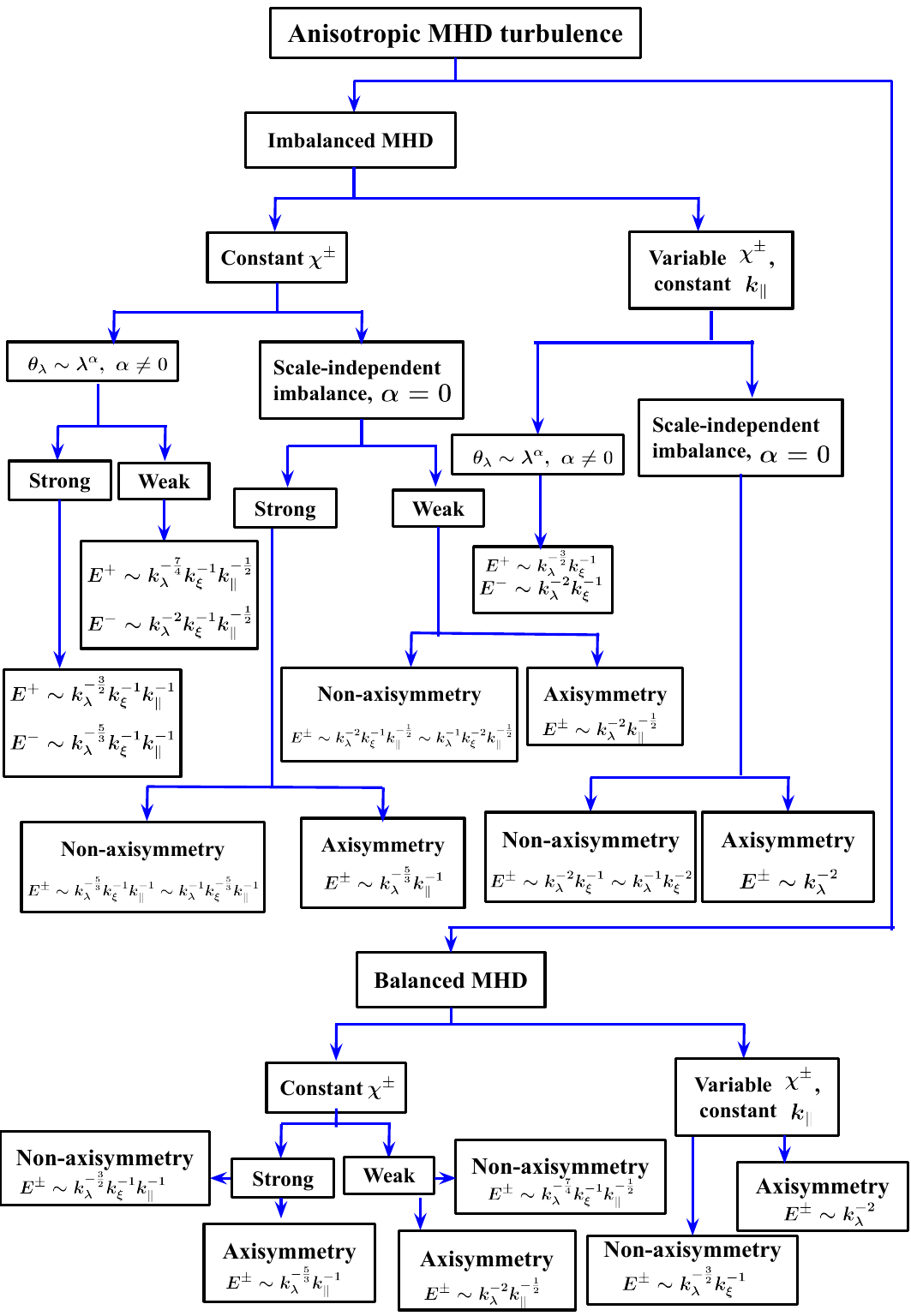}
    \caption{Schematic representation of turbulent power spectra in anisotropic MHD turbulence}
    \label{fig2}
\end{figure*}

\section{Discussion}\label{discussion}

In this paper, we have presented a general theoretical scaffold of algebraic relations between the power-law indices of parallel and perpendicular power spectra of anisotropic MHD turbulence (see figure (\ref{fig2})). For strongly imbalanced and balanced MHD turbulence with high $\bv$-$\bb$ alignment, we have derived the aforementioned relations both for strong and weak turbulence regimes. In the context of Boldyrev's phenomenology (based on scale-dependent alignment), the derived relation is consistent with non-axisymmetric perpendicular spectra $k^{-5/3}$ and $k^{-3/2}$ along and transverse to the direction of alignment, respectively. On the other hand, a scale-independent alignment leads to a $k^{-5/3}$ spectra along all directions perpendicular to the background magnetic field. For the particular case of axisymmetry, we have successfully recovered the general relation derived previously \citep{galtier2005spectral}. 

In order to provide a realistic framework for weak Alfv\'enic turbulence and it's transition to strong regime, we derived a general relation allowing $\chi^\pm$ to  vary across scales. For the case of weak turbulence, a scale-dependent alignment leads to $-3/2$ and $-2$ perpendicular spectra for the transfer of $(z^+)^2$ and $(z^-)^2$, respectively, whereas for scale-independent alignment and axisymmetry, a perpendicular $k^{-2}$ spectra is observed corresponding to both the Els\"asser variables. Note that similar power law indices do not rule out the possibility of power anisotropy between $(z^+)^2$ and $(z^-)^2$, which is expected in the case of imbalanced MHD turbulence. We also derived algebraic constraints between the perpendicular power laws of $(z^+)^2$ and $(z^-)^2$. For the strong imbalanced case, the constraint is given by $n_\lambda^++n_\lambda^-=7/2$ whereas for the balanced case, $n_\lambda^+=n_\lambda^-=3/2$, implying $n_\lambda^++n_\lambda^-=3$. This is a non-axisymmetric generalization of $n^++n^-=4$ which was previously derived for weak turbulence \citep{galtier2000}. Finally, at the transition form weak to strong turbulence, $\chi^\pm$ becomes practically constant and all the relations, derived in section \ref{model}, are recovered.

The current study is significant for reconciling the long-going debates concerning $k^{-3/2}$ versus $k^{-5/3}$ spectra \citep{beresnyak2006, Perez2009, beresnyak2011, perez2012}. As it is argued in the paper, a Boldyrev-like $k^{-3/2}$ spectra can only be obtained if we can prepare a highly aligned MHD turbulence associated with a scale-dependent alignment $(\theta_\lambda\sim \lambda^\alpha, \,\, \alpha\neq0)$. However, if the alignment is scale-independent, a perpendicular $k^{-5/3}$ spectra is always observed. Furthermore, when $k_\parallel$ varies, a parallel $k^{-2}$  spectrum is obtained irrespective of the nature of alignment. The scope of our theoretical framework may extend beyond the Boldyrev's phenomenology and can also eventually be helpful in modeling the effect of intermittency, if there is any. Moreover, the current framework can also be aligned with a realistic picture of strongly imbalanced MHD turbulence where simultaneous weak and strong cascades correspond to the stronger ($z^+$) and the weaker Els\"asser fields ($z^-$), respectively \citep{beresnyak2008}. The derived relations are also useful to find the transient spectra where the dependence on $k_\parallel$ vanishes. In such situation,  the Eqs. (\ref{epsilon15}) and (\ref{relation3}) reduce to $4\alpha_\lambda+3\alpha_\xi=11$ and $6\alpha_\lambda+5\alpha_\xi=19$ thus leading to corresponding transient spectra  $E^+(k_\lambda)\sim k_\lambda^{-2} $ and $E^-(k_\lambda)\sim k_\lambda^{-7/3}$, respectively. In the particular case of axisymmetry, we recover the transient spectra of $k_\perp^{-7/3}$ which was previously observed in numerical simulations \citep{galtier2000}.

The current methodology can further be extended to  various types of wave turbulence, including whistler waves, inertial waves, \textit{etc.} Beyond energy cascades, similar relations can also be obtained for various types of helicity cascade \citep{banerjee2016}. Finally, the entire theoretical framework is also applicable to other complex turbulent flows in binary fluid and ferrofluids \citep{pan2022, mouraya2019, mouraya2024}.  

\section*{Acknowledgments}

We thank S\'ebastien Galtier for useful discussions. S.B. acknowledges the financial support from STC-ISRO grant (STC/PHY/2023664O). 


%

\end{document}